\documentclass[preprint2]{aastex}

\newcommand\iona[2]{#1\ensuremath{\;}{\textsc{#2}}}%

\shorttitle{On the supernova associated with GRB~101219B}
\shortauthors{Sparre et al.}

\usepackage[usenames,dvipsnames]{color}

\usepackage{url}
\usepackage{graphicx}
\begin{document}


\title{Spectroscopic evidence for SN~2010\lowercase{ma} associated with
GRB~101219B \thanks{Based on observations collected at the European
Organisation for Astronomical Research in the Southern Hemisphere, Chile, under
program 086.A-0073(B). Also based on on observations obtained at the Gemini
Observatory, which is operated by the Association of Universities for Research
in Astronomy, Inc., under a cooperative agreement with the NSF on behalf of the
Gemini partnership: the National Science Foundation (United States), the
Science and Technology Facilities Council (United Kingdom), the National
Research Council (Canada), CONICYT (Chile), the Australian Research Council
(Australia), Minist\'erio da Ci\^encia e Tecnologia (Brazil) and Ministerio de
Ciencia, Tecnolog\'ia e Innovac\'on Productiva  (Argentina)}}


\author{
M. Sparre\altaffilmark{1}, 
J. Sollerman\altaffilmark{2}, 
J.~P.~U. Fynbo\altaffilmark{1}, 
D. Malesani\altaffilmark{1}, 
P. Goldoni\altaffilmark{3, 4}, 
A. de Ugarte Postigo\altaffilmark{1}, 
S. Covino\altaffilmark{5}, 
V. D'Elia\altaffilmark{6,7}, 
H. Flores\altaffilmark{8}, 
F. Hammer\altaffilmark{8}, 
J. Hjorth\altaffilmark{1}, 
P. Jakobsson\altaffilmark{9}, 
L. Kaper\altaffilmark{10},
G. Leloudas\altaffilmark{1}, 
A.~J. Levan\altaffilmark{11}, 
B. Milvang-Jensen\altaffilmark{1}, 
S. Schulze\altaffilmark{9}, 
G. Tagliaferri\altaffilmark{5},
N.~R. Tanvir\altaffilmark{12},
D.~J. Watson\altaffilmark{1},
K. Wiersema\altaffilmark{12},
R.~A.~M.~J. Wijers\altaffilmark{10}
}


\altaffiltext{1}{Dark Cosmology Centre, Niels Bohr Institute, University of Copenhagen, Juliane Maries Vej 30, 2100 Copenhagen, Denmark}
\altaffiltext{2}{Oskar Klein Centre, Department of Astronomy, AlbaNova, Stockholm University, 106 91 Stockholm, Sweden}
\altaffiltext{3}{Laboratoire Astroparticule et Cosmologie, 10 rue A. Domon et L. Duquet, 75205 Paris Cedex 13, France}
\altaffiltext{4}{DSM/IRFU/Service d'Astrophysique, CEA/Saclay, 91191, Gif-sur-Yvette, France}
\altaffiltext{5}{INAF, Osservatorio Astronomico di Brera, via E. Bianchi 46, 23807 Merate (LC), Italy}
\altaffiltext{6}{INAF, Osservatorio Astronomico di Roma, Via Frascati 33, I-00040 Monteporzio Catone, Italy}
\altaffiltext{7}{ASI, Science Data Center, via Galileo Galilei, I-00044 Frascati, Italy }
\altaffiltext{8}{Laboratoire GEPI, Observatoire de Paris, CNRS-UMR8111, Univ. Paris-Diderot 5 place Jules Janssen, 92195 Meudon France}
\altaffiltext{9}{Centre for Astrophysics and Cosmology, Science Institute, University of Iceland, Dunhagi 5, IS-107 Reykjavik, Iceland}
\altaffiltext{10}{Astronomical Institute ``Anton Pannekoek'', University of Amsterdam, Science Park 904, 1098 XH Amsterdam, The Netherlands}
\altaffiltext{11}{Department of Physics, University of Warwick, Coventry, CV4 7AL, UK}
\altaffiltext{12}{Department of Physics and Astronomy, University of Leicester, University Road, Leicester, LE1 7RH, UK}


\begin{abstract}

We report on the spectroscopic detection of supernova SN 2010ma associated with
the long gamma-ray burst GRB~101219B. We observed the optical counterpart of the
GRB
on three nights with the X-shooter spectrograph at the VLT. From weak absorption
lines, we measure a redshift of $z = 0.55$. The first epoch UV--near-infrared 
afterglow spectrum,  taken 11.6 hr after the burst, is well fit by a power
law consistent with the slope of the X-ray spectrum. The second and third epoch
spectra (obtained 16.4 and 36.7 days after the burst), 
however, display clear bumps closely resembling those of the broad-lined type-Ic
SN~1998bw if placed at $z=0.55$. Apart from demonstrating that spectroscopic SN
signatures can be observed for GRBs at these large distances,
our discovery makes a step forward in establishing a general connection between
GRBs and SNe. In fact, unlike most previous unambiguous GRB-associated SNe,
GRB~101219B has a large gamma-ray energy ($E_{\rm iso}= 4.2 \times 10^{51}$~erg), a bright afterglow, and obeys the ``Amati'' relation, thus being fully
consistent with the cosmological population of GRBs.

\end{abstract}


\section{Introduction} 

Within the last 13 years a connection between two of the most energetic
phenomena in our universe, long-duration gamma-ray bursts (GRBs, $T_{90} > 2$~s;
\citealt{1993ApJ...413L.101K}) and core-collapse supernovae (SNe), has been
established \citep[e.g.,][]{2006ARA&A..44..507W}. 

The first SN associated with a GRB was reported by \citet{1998Natur.395..670G},
who found  the extremely luminous SN~1998bw to be associated with GRB~980425.  
Subsequently, several other GRB--SNe were found, and a few of these have also
been spectroscopically confirmed, most prominently GRB~030329
\citep{2003Natur.423..847H,2003ApJ...591L..17S}, GRB~031203
\citep{2004ApJ...609L...5M},  GRB~060218
\citep{2006Natur.442.1011P,2006A&A...454..503S}, and  GRB~100316D
\citep{2011MNRAS.411.2792S,2011AN....332..262B,2010arXiv1004.2262C,2011arXiv1104.5141C}. 
All these SNe are broad-lined type Ic; they lack lines from hydrogen and helium
in their spectra and have broader lines than typical core-collapse SNe. 

Most of these spectroscopic SN-associated GRBs are low-luminosity GRBs for which the emitted energies in
the $\gamma$-ray band are in the range $10^{48}\mbox{--}10^{50}$ erg
\citep[e.g.,][]{2007A&A...463..913A,2009ApJ...701..824N}, which is 2--4 orders
of magnitude smaller than for typical GRBs detected at larger distances. They
also had comparably fainter or undetected optical and X-ray afterglows, and did
not always obey the $E_{\rm peak}$--$E_{\rm iso}$ correlation
\citep{2007A&A...463..913A}. The only well-known exception is GRB~030329 at
$z = 0.167$.
It is therefore crucial to test whether the SN-GRB connection holds in general
for high-redshift, high-luminosity GRBs. To date, evidence in this direction is
provided by the photometric detection of SN-like bumps in the late-time light
curves of several GRBs
\citep[e.g.,][]{2004ApJ...609..952Z,2004ApJ...609..962F,2010ApJ...725..625T,MNR:MNR18164}.
Spectroscopy was also obtained in a few cases 
\citep[e.g.,][]{2003ApJ...582..924G,2003A&A...406L..33D,2006ApJ...642L.103D,2008CBET.1602....1D}, 
although the contamination from the host galaxy and the faintness of the targets
makes it difficult to reach firm conclusions in some of these cases.

GRB~101219B was detected by the {\it Swift} Burst Alert Telescope (BAT) on 2010
December 19 at 16:27:53 UT. Its afterglow was promptly detected by {\it Swift}
both in the X-ray and UV/optical bands \citep{Gelbord0}. The GRB was observed to
have a duration of $T_{90} = 34 \pm 4$ s in the 15--150 keV band 
\citep{Cummings...GCN}. It was also observed by {\it Fermi}/GBM, where a
duration of  $T_{90} = 51 \pm 2$ s was determined in the 10--1000 keV band 
\citep{Horst...GCN}. The spectrum of the prompt emission as measured by GBM is
well fitted by the Band model with a peak energy at  $70\pm8$ keV. The fluence
measured by GBM is $(5.5 \pm 0.4) \times 10^{-6}$ erg cm$^{-2}$ (10--1000 keV).
Given a redshift of 0.55 (see \S~\ref{sect:z}), this corresponds to an isotropic
equivalent energy $E_{\rm iso} = 4.2\times10^{51}$ erg.

In this \textit{Letter} we report on the spectrosopic detection of a SN
associated with GRB~101219B. In \S~2 we describe our observations, in \S~3 we
present the redshift measurement and the detection of the SN. Finally, in \S~4
we offer a discussion of how this GRB-SN fits in our current understanding of
the GRB-SN connection.

For the cosmological calculations we assume a $\Lambda$CDM-universe with
$h_0=0.71$, $\Omega_\mathrm{m} = 0.27$, and $\Omega_\Lambda = 0.73$.

\section{Observations and data reduction}

We observed the afterglow of GRB~101219B using the X-shooter spectrograph \citep{2006SPIE.6269E..98D,2010SPIE.7735E..50V}
mounted at UT2 of the ESO-VLT on Cerro Paranal. X-shooter is an
echelle spectrograph with three arms covering the full spectral range from the
atmospheric cut-off around 3000~\AA{} to the $K$ band (24,800~\AA).  

An overview of the observations is given in Table~\ref{table:Obs}. For the UVB,
VIS and NIR arms, slit-widths of 1\farcs0, 0\farcs9, and 0\farcs9, respectively,
were used. The binning was $1\times2$ in all UVB and VIS exposures, and NIR-exposures were unbinned. The slit was always
aligned along the parallactic angle, and  the instrument has an atmospheric
dispersion corrector for the UVB and VIS arms.

\begin{table*}
\caption{Overview of the X-shooter observations. The $R$-band magnitudes were derived from the acquisition images.}
\label{table:Obs}
\centering 
\begin{tabular}{cccc} 
\hline\hline 
                   & Epoch 1        & Epoch 2        & Epoch 3    \\ 
\hline 
Mean time (UT)     & 2010 Dec 20.17 & 2011 Jan 5.09  & 2011 Jan 25.55 \\
Time since GRB     & 11.6 hr        & 16.4 days      & 36.9 days  \\
Exposure time (s)  & 4800           & 7200           & 7200       \\
Seeing ($\arcsec$) & 0.93--1.11     & 0.78--0.85     & 0.89--0.97 \\
Airmass            & 1.56--2.60     & 1.19--1.99     & 1.43--2.11 \\
$R$ magnitude      & $19.8\pm 0.2$  & $22.7 \pm 0.2$ & $> 22.7$   \\
\hline\hline
\end{tabular}
\end{table*}

We processed the spectra using the X-shooter data reduction pipeline
\citep{2006SPIE.6269E..80G,2010SPIE.7737E..56M} version 1.2.2. The pipeline performs all the standard 
reductions required to obtain flux calibrated echelle spectra.

All three spectra were corrected for Galactic extinction using  $R_V=3.1$ and
the prescription from \citet{Card89} with $E(B-V) = 0.02$ mag
\citep{1998ApJ...500..525S}. The spectra were reduced in staring mode and
flux-calibrated using observations of the spectrophotometric standard star LTT
3218 \citep{1993AJ....106.2392H}. We performed photometry of the transient on
the $R$-band acquisition images of X-shooter (Table~\ref{table:Obs}). Although
our photometry could be calibrated only relative to two faint USNO stars, our
results are fully consistent with nearly simultaneous measurements obtained by
GROND \citep{Olivares...GCN}. We therefore decided to scale our spectra in order
to match the GROND photometry. A relatively large correction (a factor of 2.0)
was needed for the first-epoch spectrum, probably due to the high airmass of the
observartion. The second  spectrum was also scaled to the GROND photometry
\citep{Olivares...GCN2}, but only a small correction (5\%) was necessary.

For the third epoch, we compared our spectrum with late-time $ugri$ imaging
secured by us with the Gemini South telescope equipped with GMOS on Jan 29
($u$) and on Jan 30 ($gri$), about 40 days after the BAT trigger. The flux in
the spectrum matches well our photometric measurements.

\section{Results}

\subsection{Redshift measurement}\label{sect:z}

The first epoch spectrum is characterized by a power-law spectral shape typical
for GRB afterglows. Only weak absorption lines are present, which makes it
difficult to firmly determine the spectroscopic redshift.  We tentatively
reported a redshift of 
$z = 0.5519$ 
\citep{dUP...GCN} mainly based on the \iona{Mg}{ii} 2796, 2803~\AA{} doublet. We
do not detect significant absorption at the positions of the \iona{Fe}{ii} lines
at 2382.76, 2586.65 and 2600.17~\AA, but these are located in noisier parts of
the spectrum. There is also a feature consistent with \iona{Mg}{i} 2853~\AA{} at
this redshift.

\begin{figure}
\plotone{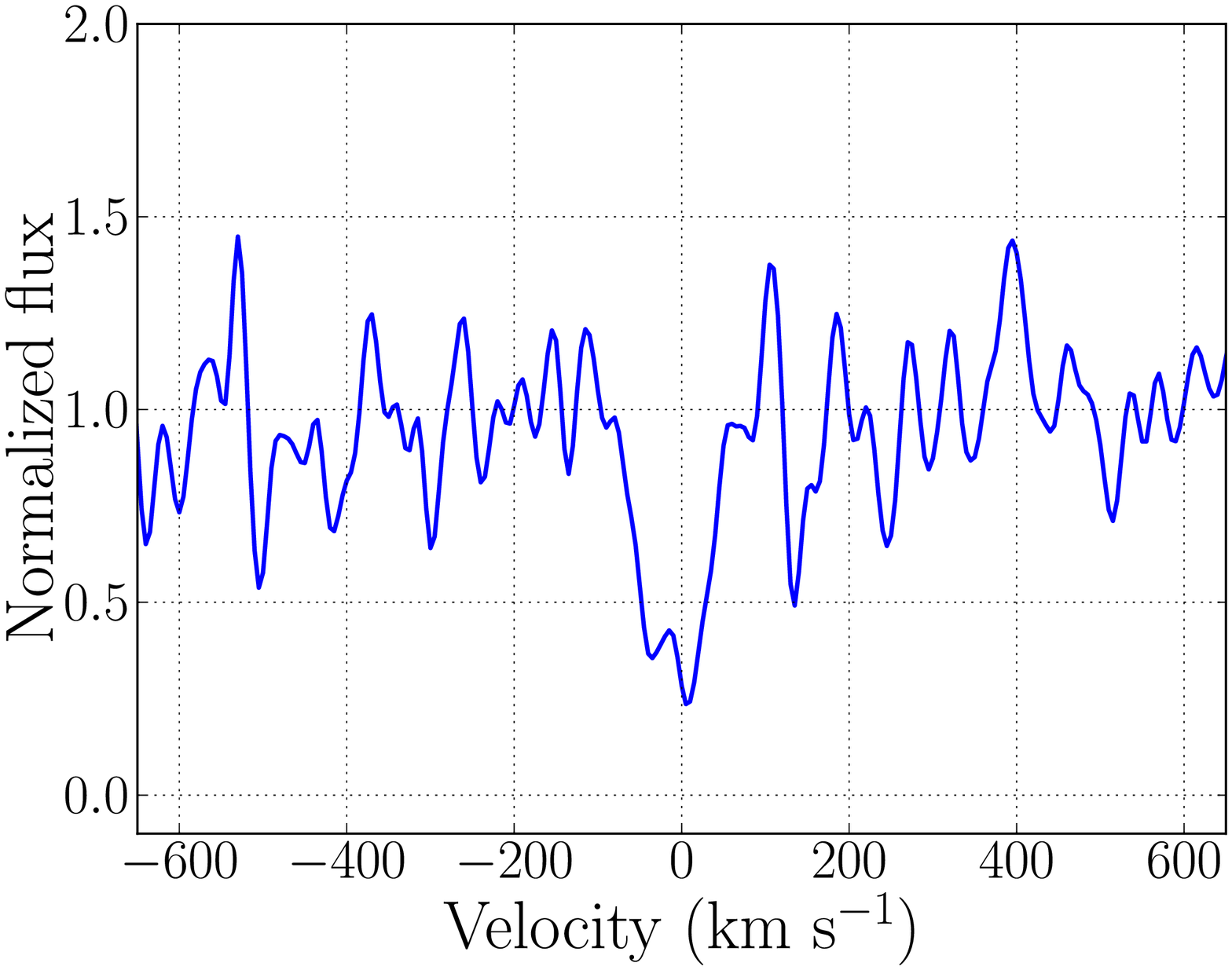}
\caption{The mean velocity profile for the first epoch spectrum, 
for three Mg absorption lines detected at
$4339.6$, $4350.7$, and $4427.5$ \AA{} assuming a redshift of $z=0.55185$. 
Each line was converted to rest frame vacuum wavelength zero
velocity before stacking.}
\label{vel}
\end{figure}

We have remeasured the redshift from the fully calibrated spectrum and find $z =
0.55185 \pm 0.00005$. To quantify the robustness of this detection we created a
mean absorption profile by stacking the positions of the three Mg lines assuming
a redshift of $z= 0.55185$. The resulting mean profile is shown in velocity
space in Fig.~\ref{vel}. There is indeed a fairly clear signal in the mean
profile assuming this redshift. The formal significance of the stacked line is
$\sim7\sigma$.  The two later epochs contain insufficient flux to significantly
improve on this estimate. We note that the redshift is consistent with the limit
of $z < 1.5$ reported from {\it Swift}-UVOT detections in the UVW2 filter
\citep{Kuin...GCN}.

\subsection{The first epoch afterglow}\label{ModellingSection}

Our first epoch afterglow spectrum is well fitted using a simple power-law
($F_\nu \propto \nu^{-\beta}$, i.e. $F_\lambda \propto \lambda^{\beta-2}$).
Using the UV -- optical range, we find $\beta_\mathrm{opt} = 0.92$
(Fig.~\ref{afterglow}), with systematic errors from the wavelength-dependence of
the slit-loss around 10\% indicated by the shaded area in the figure. The
observed slope is consistent with the X-ray power-law slope measured by the {\it
Swift} X-Ray Telescope (XRT), $\beta_\mathrm{X}  = 0.88 \pm 0.09$ \citep{evans}.

Also shown in Fig.~\ref{afterglow} are the nearly simultaneous measurements from
GROND and those from {\it Swift}-UVOT,
(\citealt{Kuin2...GCN}; Kuin, private communication). 
A temporal decay $F_\lambda
\propto t^{-1}$ is adopted to correct for the (small) time difference.

From the X-ray and optical fluxes, we calculate the broad-band spectral index
($\beta_\mathrm{OX}$) \citep[see][]{2004ApJ...617L..21J}. We find
$\beta_\mathrm{OX} = 1.02\pm 0.10$, where the error arises from the uncertainty
in the spectral index reported in the XRT spectrum and from the choice of the
two wavelength points.

The fact that $\beta_\mathrm{X}$, $\beta_\mathrm{opt}$ and $\beta_\mathrm{OX}$
are identical within their uncertainties shows that not only the slopes, but
also the normalizations of the optical and X-ray spectra are consistent, hence
both components belong to the same power-law segment. This also indicates that
little dust can be present along the line of sight (see \S~\ref{sect:dust}).

\subsection{Detection of a supernova}

\begin{figure}
\plotone{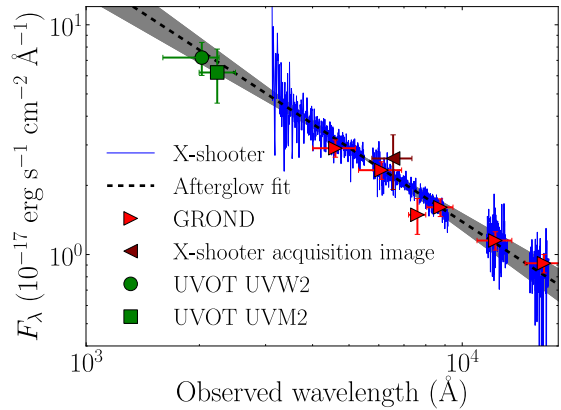}
\caption{The first epoch spectrum, which was taken 11.6 hr after the burst, 
fitted with a power-law. Also shown are fluxes from imaging with GROND, {\it
Swift}-UVOT, and our own acquisition image. The GROND-fluxes were used to fix
the normalization of our spectrum. Regions dominated by atmospheric features and
instrumental artifacts are excluded from the plot and from the fit. In the fit
we excluded the NIR ($\lambda > 11,000$~\AA), due to a possible systematic error
in the offset between the flux-calibrated spectra in the VIS and in the NIR. 
\label{afterglow}}
\end{figure}

\begin{figure*}[b]
\includegraphics[width=\linewidth]{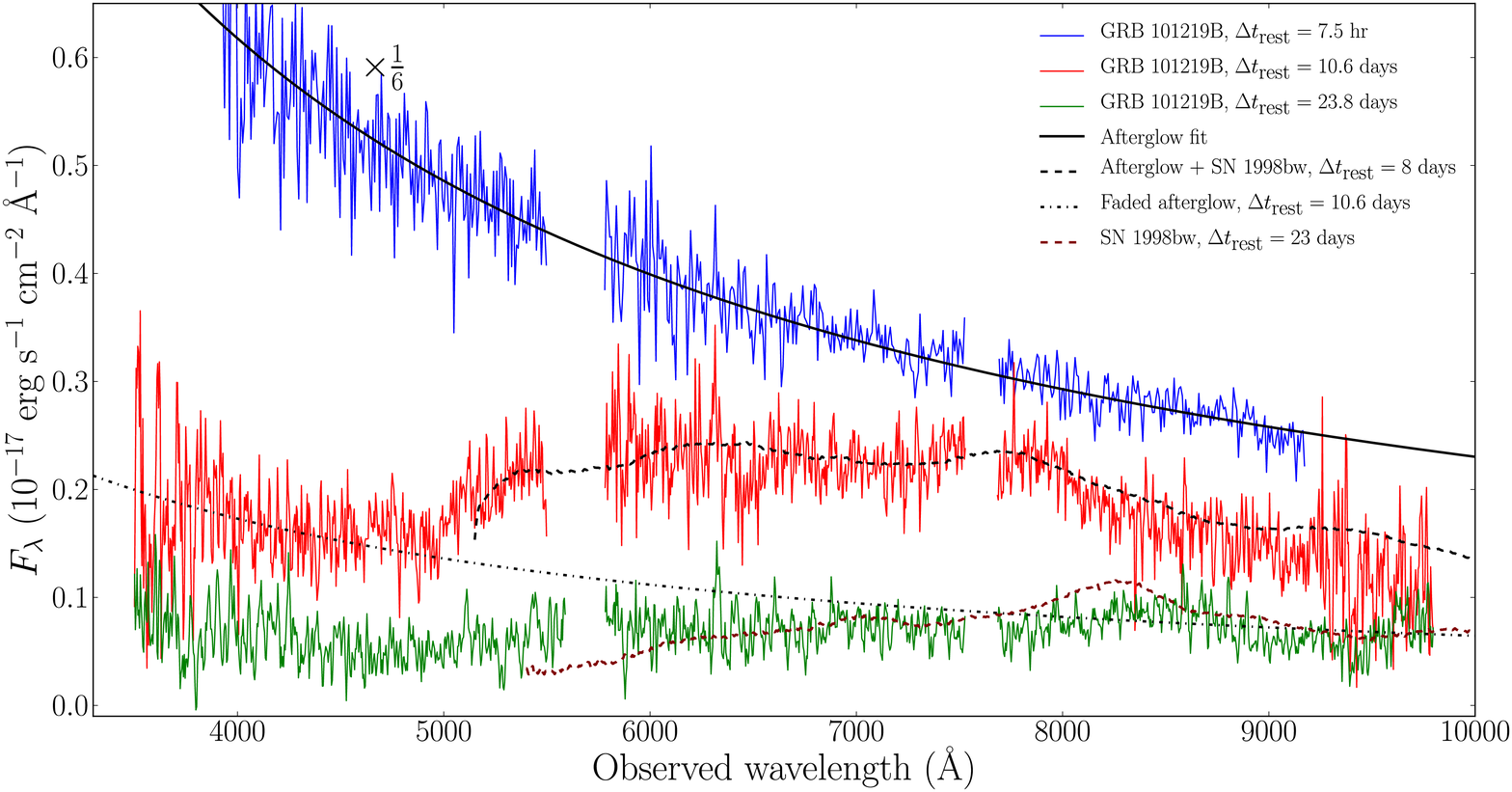}
\caption{Three epochs of spectral observations. The second epoch is compared
with a faded afterglow component and the flux from SN~1998bw (8 days after the
explosion), multiplied by 1.4. In the third epoch the contributions from the
faded afterglow and possible host galaxies are ignored, and the observations are
compared to SN~1998bw alone. The flux of the first epoch has been divided by 6
for presentation purposes. The NIR spectra are not shown, because of their low
signal-to-noise ratio in the second and the third epoch. Regions dominated by
atmospheric and instrumental features are removed from the plot. The spectra are
rebinned to a resolution of 0.8~\AA.
\label{supernova}}
\end{figure*}

In Fig.~\ref{supernova}, spectra from all the three epochs are shown together
with the first epoch power-law fit. GRB afterglows have power-law spectra, which
is seen in the first epoch  (as highlighted in Fig.~\ref{afterglow}).
This is clearly not what we observe at later times; in the second epoch spectrum
there is a prominent bump at 7800~\AA{} and an increase in flux from 5000 to
5500~\AA. The third-epoch spectrum also reveals a bump at  $\sim$8400~\AA.

We have overplotted on the second epoch spectrum a combination of the fading
power-law afterglow ($\sim20$ times dimmer than in the first epoch) and the
spectrum from SN~1998bw \citep[8 day past explosion and corrected for Galactic
extinction;][]{2001ApJ...555..900P} redshifted to $z = 0.55185$. We scaled the
flux of SN~1998bw by a factor of 1.4 to match our observations from 5000 to
8000~\AA. This reproduces fairly well both the bump and the flux increase. Epoch
2 corresponds to 10.6 days in the rest frame of the burst, for which the
SN~1998bw spectrum at 8 days is the closest available match with sufficient
wavelength coverage. We note that SN~1998bw brightened by about 0.3 mag between
8 and 10.6 days \citep{1998Natur.395..670G}, hence the flux level is consistent
between the two events.

For the third epoch we overplotted a spectrum of SN~1998bw at an epoch of 23
days after the burst in the rest frame. This spectrum of SN~1998bw was chosen 
to best match the rest-frame epoch for our spectral observations of GRB~101219B
and moved to $z=0.55185$ with no additional flux scaling.  The data are again
consistent with the presence of a SN similar to SN~1998bw, although  the
possible contribution from residual afterglow and from the host galaxy (see
\S~\ref{sect:dust}) limit the scope of this comparison.

We emphasize that in making these comparisons we have made very few  assumptions
and used few free parameters. The redshift was fixed to that  measured from the
absorption lines and the afterglow spectral slope is consistent with the X-ray
slope. The SN~1998bw spectra were chosen for epochs dictated by our 
observations.  We attribute the bluest part of the second epoch spectrum to the
optical afterglow, since the SN is unlikely to contribute substantially in this
regime \citep[due to UV line blanketing, see e.g.;][]{2003ApJ...599L..95M}. The
afterglow is thus $21 \pm 1$ times fainter than in the first epoch. The only
free parameter in the comparison with the epoch 2 spectrum is that we scaled 
the spectrum of SN~1998bw up by 40\% (consistent with the time evolution of
SN~1998bw).

The only plausible explanation for the nice match is that an energetic SN was
indeed associated with GRB~101219B. We note that \citet{Olivares...GCN2}, using
GROND, reported photometric evidence for a SN component in their light curves of
GRB~101219B with an estimated redshift of $z \approx 0.4\mbox{--}0.7$.
Following our discovery, the IAU dubbed this event SN~2010ma \citep{2011CBET}.

\subsection{Dust extinction and host galaxy continuum}\label{sect:dust}

The fact that our power-law fit agrees with the first epoch  spectrum all the
way out to the bluest end, and even extrapolates into the UVOT and XRT regimes 
(Fig.~\ref{afterglow}), demonstrates that this GRB suffered from negligible
extinction. Adding an SMC-like dust extinction when fitting the afterglow did
not improve the fit. To put an upper limit to the $V$-band absorption $A_V$, we
fitted the normalization and the spectral slope to a model, where $A_V$ was
fixed to 0.1~mag. In the UVB part of our spectrum this model gave a worse fit
than the dust-free fit, and it also did not match the UVOT points, so we
conclude that $A_V < 0.1$~mag. The lack of excess X-ray absorption
\citep{Gelbord} and the weakness of the absorption features (\S~\ref{sec:Mg})
are consistent with a tenuous host galaxy medium.

From the Gemini images obtained on Jan 29 and Jan 30 we infer the following
limits and measurements of the source: $u > 24.5$, $g = 24.1\pm0.2$, $r =
23.7\pm0.2$, and $i = 23.8\pm0.2$. This corresponds to a monochromatic flux of
about $9\times 10^{-19}$ erg cm$^{-2}$~s$^{-1}$~\AA$^{-1}$ in the $r$ band,
consistent with the third epoch spectrum. It is not clear if this corresponds to
the host galaxy alone or whether there is some contribution from the afterglow
or the SN. To test how much such a contribution could affect the luminosity of
the SN in the second epoch, we assumed as an upper limit that the host galaxy
has a wavelength-independent flux corresponding to what is measured in the third
epoch spectrum. To this value we added a faded afterglow and a contribution from
SN~1998bw. In this way, the SN would be 30\% fainter than assuming no host
contamination. The host galaxy is thus unlikely to affect the conclusion that a
luminous SN component is dominating our epoch 2 spectrum.

We have also searched the spectra for the strongest host galaxy  emission lines. Unfortunately, at $z=0.55185$ [\iona{O}{ii}]~$\lambda$3727 and H$\alpha$ fall in the transition regions between the UVB/VIS and VIS/NIR arms, respectively.
However, [\iona{O}{iii}]~$\lambda$5007 is located in a clear part of the
spectrum. To derive an upper limit on the line flux we added artificial emission
lines of increasing strength to the data, until the line was easily detectable.
For [\iona{O}{iii}]~$\lambda$5007 we find that an emission line of
$3\times10^{-17}$ erg~s$^{-1}$~cm$^{-2}$ would have been detected. At the
redshift of the burst this corresponds to a luminosity of 
$L_{[\rm \iona{O}{iii}]} = 4 \times 10^{40}$ erg~s$^{-1}$. While this limit is
not unprecedented (for example, the host galaxy of GRB~030329 has  $L_{[\rm
\iona{O}{iii}]} = 3\times10^{40}$ erg~s$^{-1}$; Hjorth et al.\ 2003), this value
lies at the lower end of the distribution \citep{2009ApJ...691..182S},
consistent with a faint host galaxy.

\subsection{Mg absorption in the host galaxy}
\label{sec:Mg}

Table~\ref{table:EW} displays the equivalent widths of the absorption features
in the first epoch spectrum. \iona{Mg}{II}$~\lambda$2796 is stronger than
\iona{Mg}{II}$~\lambda$2803
and their ratio ($1.37 \pm 0.37$) is consistent with the average value ($1.16
\pm 0.03$) of the composite spectrum presented by \citet{2011ApJ...727...73C}.
These lines are weaker than the ones of the typical GRB spectrum, but similar or
even weaker cases have been seen before (e.g., GRB~050922C:
\citealt{2008A&A...492..775P}; GRB~070125: \citealt{2009ApJS..185..526F}). The
detection limits of iron lines are consistent with the weak magnesium features.

\begin{table}
\caption{Equivalent widths for the absorption features in the first epoch
spectrum, both in the observer and rest frames. Limits are at 3$\sigma$.}
\label{table:EW}
\centering 
\begin{tabular}{ccc}
\hline\hline 
Feature                      & EW$_\mathrm{obs}$ (\AA) & EW$_\mathrm{rest}$ (\AA) \\ \hline 
\iona{Fe}{II} $\lambda$2586  & $<$0.96                 & $<$0.62       \\
\iona{Fe}{II} $\lambda$2600  & $<$0.90                 & $<$0.58       \\
\iona{Mg}{II} $\lambda$2796  & 1.34$\pm$0.25           & 0.86$\pm$0.16 \\
\iona{Mg}{II} $\lambda$2803  & 0.98$\pm$0.20           & 0.63$\pm$0.13 \\
\iona{Mg}{I}  $\lambda$2852  & 0.89$\pm$0.18           & 0.57$\pm$0.12 \\
\hline\hline
\end{tabular}
\end{table}

\section{Discussion}

In this paper we report on our detection of the spectral signature of a SN,
SN~2010ma, in the fading afterglow spectrum of GRB~101219B.

The redshift was measured to be $z = 0.55185$ based on our detections of weak
\iona{Mg}{II} absorption lines in the first epoch spectrum. At this redshift, a
SN similar to SN~1998bw can well match our second epoch spectrum. The broad
spectral features indicate high velocities in the expanding ejecta, whereas a
normal SN Ib/c clearly does not match the observations. A SN interpretation is
consistent with the drop at $\sim 5000$~\AA{} being due to UV blanketing in the
SN spectra, where the UV part is instead dominated by the blue afterglow
component. The host galaxy appears to be relatively faint, which likely
contributed to obtain a clean signal from the supernova. There is little room
for host galaxy extinction.

The GRB itself is clearly of long duration and has, at the measured redshift, an
isotropic equivalent energy of $4.2\times10^{51}$ erg. It has a bright afterglow
with a standard light curve, and obeys the $E_{\rm p} \mbox{--} E_{\rm iso}$
(``Amati'') relation \citep{2007A&A...463..913A}. It is thus fully
representative of the high-luminosity, routinely observed high-redshift
population of GRBs, for which we now can provide a robust, spectroscopic
association with a SN.

Since 4 out of the 5 previously unambiguously spectroscopically confirmed
SN--GRBs were in fact rather low luminosity bursts with unusual afterglows and
spectral properties, there have been suggestions in the literature
\citep[e.g.,][]{2007ApJ...654..385K} that the GRB--SN association is only proven
for the local universe. Our new association favours a GRB-SN connection also for
the main population of cosmological GRBs. This confirms previous observations of
photometric bumps and tentative spectroscopic SN detections for distant bursts.
In the recent compilation by \citet{2011arXiv1104.2274H}, the evidence for the
SN association of GRB 101219B was graded just after the five `ironclad' cases
mentioned in \S~1. The fact that we were able to see the SN component emerge
over three epochs with the same instrumentation leaves little room for confusion
by afterglow or host galaxy emission --- also thanks to the faintness of the
host. The X-shooter spectrograph will with no doubt continue to contribute to this
area of research.

\section*{Acknowledgements}
We acknowledge Michael Andersen, Massimo Della Valle, Suzanne Foley, Paul Kuin,
and Paul Vreeswijk for helpful comments. The Dark Cosmology Centre is funded by
the Danish National Research Foundation. GL is supported by a grant from the
Carlsberg foundation. RAMJW acknowledges support from the European Research
Council via Advanced Investigator Grant no. 247295. SS acknowledges support by a
Grant of Excellence from the Icelandic Research Fund.\vspace*{0.1cm}

\bibliographystyle{apj}

\end{document}